\documentclass[prd,twocolumn,preprintnumbers,nofootinbib]{revtex4}
\usepackage{graphicx}
\usepackage{color}
\usepackage{float}
\newcommand{\rthis}[1]{\textcolor{black}{#1}}
\usepackage{amsfonts,amsmath,amssymb}
\usepackage[plainpages=false, colorlinks=true, anchorcolor=blue, linkcolor=blue, citecolor=blue, bookmarks=false]{hyperref}
\usepackage{natbib}
\usepackage{enumitem}
\usepackage{array}
\pdfoutput=1
\begin{document}
\newcommand{\bthis}[1]{\textcolor{black}{#1}}
\newcommand{\apjl}{Astrophys. J. Lett.}
\newcommand{\apjs}{Astrophys. J. Suppl. Ser.}
\newcommand{\aap}{Astron. \& Astrophys.}
\newcommand{\aj}{Astron. J.}
\newcommand{\araa}{Ann. Rev. Astron. Astrophys. } 
\newcommand{\mnras}{Mon. Not. R. Astron. Soc.}
\newcommand{\ssr}{Space Science Revs.}
\newcommand{\apss}{Astrophysics \& Space Sciences}
\newcommand{\jcap}{JCAP}
\newcommand{\pasj}{PASJ}
\newcommand{\pasp}{PASP}
\newcommand{\pasa}{Pub. Astro. Soc. Aust.}
\newcommand{\physrep}{Phys. Rep.}
\title{A test of   galaxy cluster fundamental plane for  the X-COP sample}
\author{S. \surname{Pradyumna}}\altaffiliation{E-mail:ep18btech11015@iith.ac.in}

\author{Shantanu  \surname{Desai}}  
\altaffiliation{E-mail: shntn05@gmail.com}

\begin{abstract}
We test the galaxy cluster fundamental plane   using the X-COP sample of 12 clusters. The fundamental plane is given by the relation $T_X \propto M_s^{\alpha} r_s^{\beta}$, where $T_X$, $M_s$, and $r_s$ correspond to the gas temperature, NFW  halo mass, and scale radius, respectively. We did this analysis using two different temperatures: the error-weighted temperature in $(50-500)h^{-1}$ kpc as well as the mass-weighted temperature in the same range.
With both these temperatures, we find a very tight fundamental plane  with dispersion of about 0.02 dex. The best-fit values for $\alpha$ and $\beta$ are in-between those expected from virial equilibrium and self-similarity  solution for secondary infall and collapse, with $\alpha$ being closer to the virial expectation. 
Our best-fit values are also consistent with a recent re-analyses of the fundamental plane for the CLASH sample, after excluding the hottest clusters.

\end{abstract}
\affiliation{Department of Physics, Indian Institute of Technology, Hyderabad, Telangana-502285, India}
\maketitle

\section{Introduction}

Galaxy clusters are the biggest gravitationally collapsed  objects  in the universe and have proved to be wonderful laboratories for Cosmology, galaxy evolution, and fundamental Physics~\cite{Vikhlininrev,Allen,Borgani,Desai18,Bora,Borac}.  Although galaxy cluster along with CMB  observations  are broadly consistent with the standard $\Lambda$CDM model consisting of 70\% dark energy and 25\% dark matter~\cite{Planck18}, there are still some  vexing issues with this standard cosmological model. Some of these problems with the standard model include the core-cusp and the  missing satellites problem~\cite{Bullock}, Hubble 
constant tension~\cite{Divalentino}, $\sigma_8$  tensions between cluster and CMB~\cite{Benisty}, failure to detect cold dark matter candidates in laboratory based experiments~\cite{Merritt}, Lithium-7 problem in Big-Bang nucleosynthesis~\cite{Fields}, radial acceleration relation in spiral galaxies with very low scatter~\cite{McGaugh16}, constant dark matter halo surface density~\cite{Donato},  CMB anomalies at large angular scales~\cite{Copi},  etc. An up-to-date  summary of the  challenges for $\Lambda$CDM model can be found in ~\citet{Periv}.  Therefore a large number of alternatives  to the standard cosmological model  have been proposed to account for some of these anomalies~\cite{alternatives,Banik}. Galaxy clusters have proved to be very good probes to test some of the above  anomalies observed on galactic scales such as radial acceleration relation or constant surface density, which could break the degeneracy between alternatives to $\Lambda$CDM and more complicated baryonic feedback models~\cite{Chan14,Tian,Pradyumna20,Pradyumna21,Gopika,Gopika21}. Therefore,  clusters continue to remain as flagship objects to probe the various anomalies observed in $\Lambda$CDM.

In 2018 ~\citet{Fujita2018} (F18, hereafter) found a tight  empirical correlation between the characteristic radius, characteristic dark matter halo mass, and temperature for a sample of 20   galaxy clusters from the CLASH survey in the redshift range between 0.186-0.686~\cite{Postman12}, which has been dubbed  as the fundamental plane (FP).

To see where the FP comes, from we first note that the mass $M(r)$ of the NFW dark matter profile~\cite{NFW} at a radius $r$ with halo concentration ($c' \equiv \frac{r}{r_s}$)\footnote{We have used the symbol $c'$  (instead of the normally used symbol $c$) for  the halo concentration to distinguish it from $c$ in Eq.~\ref{eq:FP}}   is given by~\cite{Diemer} 
\begin{equation}
    M(r) = 4 \pi \rho_s r_s^3 \left[\ln(1+c') - \frac{c'}{1+c'} \right],
    \label{eq:NFWmass}
\end{equation}
where $r_s$ is the scale radius for which the logarithm of the slope of the density is equal to -2;  and $\rho_s$ is the characteristic density which change from halo to halo~\cite{NFW,Garcia2021} \rthis{and  is given by:}
\begin{equation}
    \rho_s = \frac{200}{3} \frac{\rho_{cz} {c'}^3_{200}}{\left[\ln(1+c_{200}') -\frac{c_{200}'}{1+c_{200}'}\right]},
\label{eq:rhos}
\end{equation}
\rthis{where $\rho_{cz}$ is the critical density for an Einstein-Desitter universe and is given by $\rho_{cz} = \frac{3H^2(z)}{8 \pi G}$.}
From Eq.~\ref{eq:NFWmass}, one can define a mass $M_s$ enclosed in a sphere of radius $r_s$ as:
\begin{equation}
    M_s= 4\pi \rho_s r_s^3 [\ln(2)-0.5]
    \label{eq:ms}
\end{equation}

The  FP is characterized by a linear relation  between the logarithms of $r_s$, $M_s$ (Eq.~\ref{eq:ms}), and $T_X$, characterized by
\begin{equation}
    a\log r_s  + b \log M_s + c \log T_X = d
    \label{eq:FP}
\end{equation}
 where,  $T_X$ is the  X-Ray temperature. For the CLASH cluster sample,
 $r_s$ and $M_s$  were obtained  from joint weak and strong lensing observations~\cite{Umetsu16}, and $T_X$ was obtained between 50-500 kpc, after excluding the cool core at the center~\cite{Postman12}. \rthis{F18 chose a radial range of $< 500$ kpc for calculating the temperature, in order to retain a memory of the formation time of the halo, as the inner regions of clusters ($r<r_s$) should
reflect their rapid growth early in their formation.}

Alternately, the  FP can also be recast as a regression relation between $T_X$, $r_s$, and $M_s$, given by 
$T_X \propto M_s^{-b/c} r_s^{-a/c}$.
For clusters in virial equilibrium, one expects $T_X \propto \frac{M_{vir}}{r_{vir}}$ \rthis{from the virial theorem~\cite{Voit05}. The quantity $\frac{M_{vir}}{r_{vir}}$ is proportional to $M_s/r_s$  for a NFW profile  because of the presence of a characteristic radius  ($r_s$).} Therefore, in virial equilibrium, we get $T_X \propto \frac{M_s}{r_s}$ or equivalently $b/c=-1$ and $a/c= 1$.
The best-fit values obtained by F18 for the CLASH sample are given by  $a= 0.76^{+0.03}_{-0.05}$, $b = -0.56 \pm 0.02$, and $c=0.32 \pm 0.1$, or equivalently $T_X \propto M_s^{1.8 \pm 0.5} r_s^{-2.3 \pm 0.7}$. The scatter around the FP was found to be about 0.045 dex.  Therefore, the F18 best-fit values are  inconsistent with that predicted from virial equilibrium.  F18 further confirmed the FP relation with a catalog of about 400 simulated clusters from the MUSIC sample~\cite{Music}, and showed that the  observed FP is consistent with the values obtained for the CLASH sample. F18 then pointed out that the best-fit values of $a$, $b$, and $c$  are consistent with   Bertschinger's self-similarity solution for the secondary infall and accretion~\cite{Bertschinger85} (See also ~\cite{Delpopolo09,Fujitagalaxies}), which predicts $T_X \propto  M_s^{1.5}r_s^{-2}$. F18 also argued that the FP is a very good probe of the mass accretion history of the cluster. 
The F18 work was then followed up with a similar analysis~\cite{Fujita2} for a sample of 44 X-ray clusters~\cite{Ettori10}. This aforementioned work obtained  a best-fit FP characterized by  $a=0.71^{+0.06}_{-0.10}$, $b=-0.53^{+0.02}_{-0.01}$, $c=0.46^{+0.13}_{-0.11}$ with a scatter of 0.039 dex~\cite{Fujita2} which corresponds to $T= M_s^{1.15^{+0.29}_{-0.39}} r_s^{-1.54^{+0.51}_{-0.68}}$.   Therefore, the best-fit values were consistent with the values found using the CLASH sample. For this catalog, this relation was also separately tested for cool-core, non-cool-core, intermediate-cool-core,  and it was found that the best-fit parameters are mostly consistent across the samples~\cite{Fujita2}. Because of its tight scatter, the FP has also been proposed as a tool for cluster mass calibration~\cite{Fujitagalaxies}.

Recently, the FP  was comprehensively tested using a sample of about 250 simulated clusters~\cite{Garcia2021} (G21, hereafter) from the Three Hundred project sample~\cite{Cui18}. For their analysis, G21 used two different definitions for the temperature: mass-weighted and spectroscopic temperatures. Also the FP was tested with temperatures at three different radii: $r_{200}$, $r_{500}$, and $500 h^{-1}$ kpc. G21 also tested the evolution of the FP as a function of the  redshift and cluster dynamical state.
They confirmed the existence of the FP, but found a relation closer to the virial prediction  instead of the self-similar solution found in F18, when using the mass-weighted temperatures (cf. Table 3 of G21). When the gas temperatures were measured inside 500 $h^{-1}$ kpc, the FP parameters were found to be  in-between the virial and self-similar solution. Finally, G21 found that the clusters at $z=1$ are consistent with the virial expectations. The dispersion of the FP for the simulated sample varied from 0.015 to 0.03 dex.
Given this discrepancy with respect to  the results in F18 and the observed variations with respect to the   temperature definition used  as well as redshift, G21 reanalyzed the CLASH sample and found that the shift towards the self similar solution found by F18 is caused by four clusters with the highest X-ray temperatures ($T_X > 12$ keV). After excluding these four objects as none of the simulated clusters had $T_X>$ 12 keV, G21  found the FP to be closer to the virial expectations. 
G21 also  re-analyzed only the relaxed cluster sample in F18 (consisting of 13 objects), and  found that the fundamental plane is compatible with both the virial as well as the self-similar solution. More details about the findings in G21 can be found in ~\cite{Garcia2021}.

Therefore, given the somewhat discordant results between  F18 and G21, it is important to re-affirm if  other cluster samples follow a tight FP between  $r_s$, $M_s$, and $T_X$, and whether the best-fit values are closer to the virial or self-similar expectations. In the past three decades, a large number of X-ray based galaxy cluster surveys have mapped out large regions of the sky to discover
new galaxy clusters. There have also been many pointed X-ray followup campaigns  to image clusters discovered in  optical or SZ surveys~\cite{Desai12,Song12}. Therefore, we carry out another test of the fundamental plane using a sample of 12 clusters imaged as part of the XMM Cluster Outskirts Project (X-COP) survey~\cite{XCOPmain}.  

The outline of this manuscript is as follows.
We discuss the data sample used for our analysis in Sec.~\ref{sec:data}. Our analysis and results are discussed in Sec.~\ref{sec:results}. Our conclusions can be found  in Sec.~\ref{sec:conclusions}.
We assume standard $\Lambda$CDM model with $H_0 = 70 \mathrm{\, km\,s^{-1}\,Mpc^{-1}}, \Omega_m = 0.3$, and $\Omega_\Lambda = 0.7$. This implies $h = H_0 / (100 \mathrm{\, km\,s^{-1}\,Mpc^{-1}}) = 0.7$.

\section{Data and Methods}
\label{sec:data}
For testing the fundamental plane, both the X-Ray temperature and the estimates of the best-fit NFW parameters to  the mass profiles are required. For this work, we  have used the X-COP sample~\cite{XCOP1}, which we describe below.
\subsection{The X-COP sample}
X-COP (PI D. Eckert)  is a very large observing program on the  XMM-Newton telescope, which studies X-ray emission in the outskirts of galaxy clusters. This project observed a sample of 12 massive galaxy clusters, selected from the Planck all-sky SZ catalog~\cite{planck_2014}, with SNR$>12$, corresponding to a mass threshold of  $M_{500}>3\times 10^{14} M_{\odot}$, in the redshift range of $0.04-0.1$. These clusters have measured temperatures less than 12 keV with A2319 having the maximum measured temperature of 10.3 keV. More details about the X-COP sample can be found in \cite{XCOP1,XCOPmain}. We have previously used this sample for a test of the radial acceleration relation~\cite{Pradyumna20}.
The parameters for the NFW profiles have been obtained using the backward method~\cite{Ettori10}, where a parametric profile is posited for the dark matter. The accuracy of the backward method has been compared with other non-parametric methods, and the results agree to within 5\% at 1.5 Mpc~\cite{XCOP1}.
This profile is then combined with the gas mass density estimated from deprojected surface brightness profile to predict the temperature profile. This estimated temperature profile is then compared with the temperature profile estimated from the spectral analysis of the X-Ray data to obtain the NFW parameters. The uncertainty in the hydrostatic mass is about 5\%. The median ratio of the hydrostatic to weak lensing masses is about 88\%~\cite{XCOP1}. By comparing the gas mass fraction with the universal gas mass fraction values, the median non-thermal pressure support was found to be  about 6\%  at $R_{500}$~\cite{Eckertbias}. We use the $c_{200}$ and $M_{200}$ values \rthis{for the X-COP clusters} obtained from the backward method and tabulated in Table 1 of \cite{XCOP1} to obtain $r_s$ and $M_s$. For this sample, the temperature of the  X-Ray emitting gas is obtained from the deprojected spectral estimates. 
We carried our analyses using two definitions of the X-Ray temperatures for a given cluster to find the FP. 
Firstly, in Sec.~\ref{sec:t300}, we take  the error weighted mean of the  temperature measurements\footnote{The temperature measurements are provided in the form of fits files at \href{https://dominiqueeckert.wixsite.com/X-COP}{X-COP Website}} in $(50 - 500)h^{-1}$ kpc (this is denoted as $T_{ew}$ hereafter).  \rthis{For our choice of $h=0.7$, this corresponds to 71-714 kpc.}
We define a temperature this way to enable easy comparison with other works. A similar method of using the error weighted temperature is employed by \citet{Fujita2} in their analysis on \citet{Ettori10} X-ray sample.
Secondly, we explore the FP  using the mass weighted temperature in $(50-500)h^{-1}$ kpc region (denoted as $T_{mw}$ hereafter) in Sec.~\ref{subsec:Tmw}.

\subsection{Estimation of $r_s$ and $M_s$}
Both $r_s$ and $M_s$ can be obtained from the NFW fits.
For obtaining $r_s$ and $M_s$, we make use of $c_{200}$ and $M_{200}$ data provided for this X-COP sample. $R_{200}$ can be calculated from $M_{200}$ using
\begin{equation}
    M_{200} = \frac{100 R_{200}^3 H^2(z)}{G}
    \label{eq:M200_nfw}
\end{equation}
Where $H^2(z) = H_0^2 (\Omega_m(1+z)^3 + \Omega_\Lambda)$ and $G$ is the universal gravitational constant.
Now, $r_s$ can be calculated from $r_s = R_{200}/c_{200}$ and $\rho_s$ can be estimated from Eq.~\ref{eq:rhos}. We then plug $r_s$ and $\rho_s$ in Eq.~\ref{eq:ms} to obtain $M_s$. The uncertainties on $M_{200}$ and $c_{200}$ have been provided by \cite{XCOP1}. These are propagated onto $r_s$ and $M_s$.


\section{Analysis and Results}
\label{sec:results}

We now describe the procedure for determining the best-fit FP and our results for the X-COP sample of galaxy clusters. 

\subsection{FP using error weighted temperature}
\label{sec:t300}
We now combine $M_s$, $r_s$, and $T_{ew}$, which are estimated as detailed in the previous section.
Firstly, we normalize $(M_s, r_s, T_{ew})$ by the mean values $(M_{s,0}, r_{s,0}, T_{ew,0})$ for the X-COP sample. The mean values of $M_s, r_s$, and $T_{ew}$ for the X-COP sample are $M_{s,0} = \rthis{2.6} \times 10^{14}\mathrm{\; M_\odot}$, $r_{s,0} = \rthis{746}\mathrm{\; kpc}$  and $T_{ew, 0} = 6.3$ keV respectively. This scaling is done in order to center the data at the origin, so that the best-fit plane passes through the origin and therefore, $d$ in Eq.~\ref{eq:FP} can be set to zero. For the remaining analysis, we define:
{\small
\begin{equation}
\begin{aligned}
    x &\equiv \log_{10} \left({r_s}/{r_{s,0}} \right) \\
    y &\equiv  \log_{10} \left({M_s}/{M_{s,0}} \right)\\
    z &\equiv  \log_{10} \left( {T_{ew}}/{T_{ew,0}}\right)
\end{aligned}
\label{eq:xyz}
\end{equation}
}
This alters Eq.~\ref{eq:FP} to the form: $ax+by+cz=0$. We then use principal component analysis (PCA) to fit a plane to this data.  PCA in three dimensions gives three unit orthogonal vectors $P_1, P_2$, and $P_3$, with $P_1, P_2, P_3$ in  increasing order of the  square of the distances from the points. By this property of PCA, $P_1$ is the vector along which the data shows the maximum variance,  and $P_3$ is the vector which is normal to the best-fit plane, given by  $P_1$ and $P_2$. 

\begin{figure*}
    \centering
    \includegraphics[width=2\columnwidth]{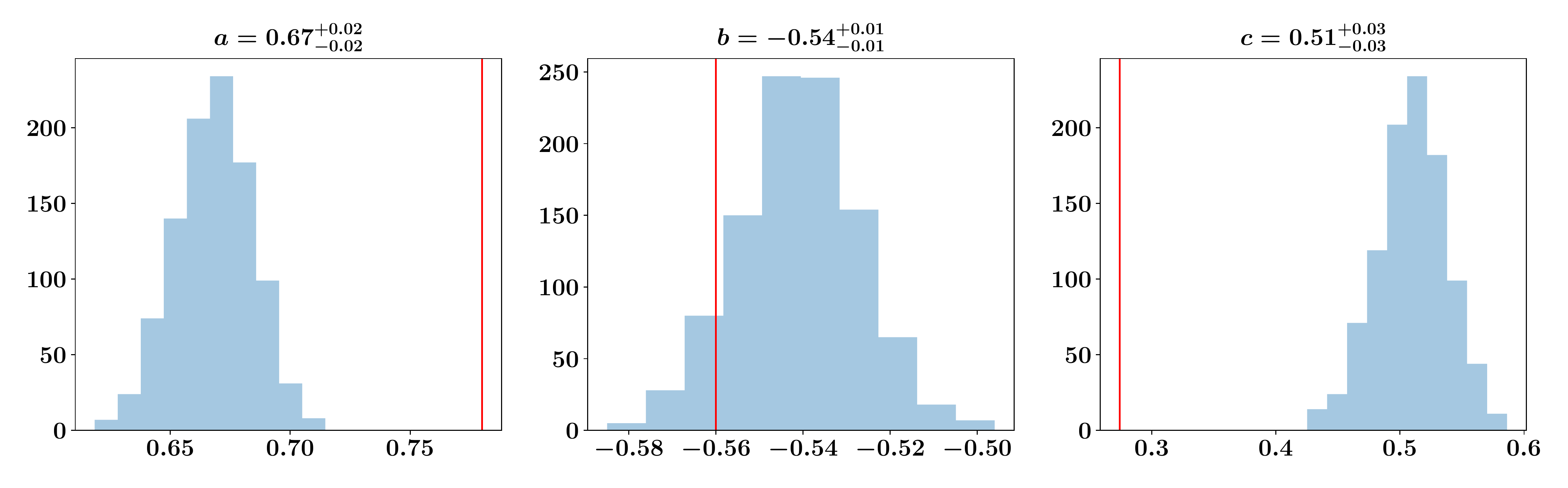}
    \caption{The distribution of the PCA-based estimates of $a, b$, and $c$  obtained by generating 1000 synthetic datasets by sampling $M_{200}, c_{200}, T_{ew}$ from Gaussian distributions with the observed measurements as the mean,  and the standard deviations from  the uncertainties. The red lines show the fit obtained by F18 for the CLASH sample~\cite{Fujita2018}.}
    \label{fig:synth}
\end{figure*}

Since PCA does not automatically incorporate the errors in the parameters, we estimate the uncertainties in the parameters using Monte Carlo simulations with 1000 realizations. For this, we randomly sample $M_{200}, c_{200}$, and $T_{ew}$ from $\mathcal{N}(M_{200}, \sigma_{M_{200}}), \mathcal{N}(c_{200}, \sigma_{c_{200}})$, and $\mathcal{N}(T_{ew}, \sigma_{T_{ew}})$ respectively, where $\mathcal{N}(A, \sigma_A)$ for a quantity $A$ denotes a Gaussian distribution with mean $A$ and standard deviation $\sigma_A$. Here, $\sigma_A$ is the estimated uncertainty of the quantity $A$.
Then, we re-estimate $M_s$ and $r_s$ as explained in the previous section. Now, PCA is applied again to each of these synthetic datasets.
The vector along which the X-COP data lies is given by $P_1 = (0.65 \pm 0.01, 0.76 \pm 0.01, -0.03\pm 0.03)$.
Further, since $P_3$ is the unit vector normal to the best-fit plane, from Eq.~\ref{eq:FP}, $P_3 = (a,b,c)$. From the  application of PCA to the aforementioned synthetic datasets, we obtain arrays of $a$, $b$, and $c$.
Fig.~\ref{fig:synth} shows the histograms so obtained for $a, b$, and $c$, with the red vertical lines being the best-fit values obtained for CLASH sample by F18. We find that $a = 0.67 \pm 0.02, b = -0.54 \pm 0.01$, and $c = 0.51 \pm 0.03$.
We can rearrange Eq.~\ref{eq:FP} as $T_{ew} \propto M_s^{-b/c}r_s^{-a/c}$, as in G21  and  define $\alpha \equiv -b/c$ and $\beta \equiv -a/c$. From the best-fit values of  $a$, $b$, $c$ , we obtain $\alpha = 1.06\pm 0.02$ and $\beta = -1.31\pm 0.12$. To calculate the scatter in dex, we followed the same prescription as in G21,  and calculated half of the difference between the $16^{th}$ and $84^{th}$ percentile of the distances from the FP given by $d_i = |ax_i + by_i + cz_i|$. With this method, we find a scatter  of 0.019 dex.

We also did an independent estimate of $\alpha$ and $\beta$ by recasting the fundamental plane as a linear regression relation between $T_{ew}$, $r_s$ and $M_s$ in logarithmic space and estimating the best-fit parameters using Bayesian regression. The results of this analysis can be found in the Appendix A and our best-fit values for $\alpha$ and $\beta$ agree with the PCA-based analysis within $1\sigma$ error bars.

We find that our value for $b$ is consistent with F18, but not $a$ and $c$. The values of $\alpha$ and $\beta$ are also discrepant with  respect to the self-similar solutions. They are in-between the virial and self-similar expectations (with $\alpha$ being closer to the virial values).The value of $\beta$ agrees with the values obtained for the  Three Hundred project clusters in $(50-500) h^{-1}$ kpc, but $\alpha$ is discrepant.  Both our FP values are consistent with the  re-analyses of the CLASH sample done in G21, after excluding the clusters with $T_X >$ 12 keV (cf. Table~\ref{tab1}).
\rthis{To test the correlation between the $a, b$, and $c$, we  determine the  Pearson correlation coefficients using the FP  parameters of the synthetic datasets.
Pearson correlation ($r$)  is an indicator of linear correlation that takes values from $-1$ to $+1$, where $-1$ corresponds to complete anti-correlation, $0$ corresponds to nonexistence of linear correlation and $+1$ corresponds to complete positive correlation~\cite{astroML}. Table~\ref{tab:pearsonr} shows the coefficients obtained for $a$, $b$, and $c$. Based on the  Pearson $r$ values, we can infer that $(c, a)$  and $(a,b)$ are anti-correlated, whereas $(b,c)$ is correlated. 
We also  applied Pearson correlation coefficient to the $\alpha$ and $\beta$ values from the MCMC chains  discussed in the Appendix. We find that $\alpha$ and $\beta$ are anti-correlated, which is in accord with the findings in G21.}

\subsection{FP with mass weighted temperatures}
\label{subsec:Tmw}
G21  found that  the FP parameters are sensitive to the  choice  of the X-Ray gas temperature. 
Given the excellent quality of the  X-COP data, we also check the sensitivity of the FP results to a different variant of the temperature. For this purpose, we consider a  mass-weighted temperature defined as:
\begin{equation}
    T_{mw}(r) = \frac{\int_0^r T(r) dM}{\int_0^r dM} = \frac{\int_0^r T(r) \rho_{gas}(r) dV}{\int_0^r \rho_{gas}(r) dV}
    \label{eq:Tmw}
\end{equation}
where $\rho_{gas}$ is the gas density as a function of radius from the center and $dV = 4\pi r^2 dr$ is the volume element. The gas density can be evaluated as $\rho_{gas} = (n_e + n_p)m_u \mu \approx 1.85 n_e m_u \mu$. Here, $n_p$ and $n_e$ are the proton and electron densities respectively and $\mu m_u \approx 0.6 \text{ a.m.u.}$ is the mean molecular weight of the X-Ray gas. The measurements of $n_e$ for X-COP sample are available from the deprojected XMM surface brightness profile and can be downloaded from  the X-COP website.
The $\rho_{gas}$ so estimated and the temperature measured are interpolated and the integrals are evaluated numerically.
We integrate the numerator and denominator of Eq.~\ref{eq:Tmw} in $(50-500)h^{-1}\mathrm{\; kpc}$ range.
Redefining temperature this way and repeating the PCA-based analysis to find the FP, gives $a = 0.68\pm 0.02$, $b = -0.56 \pm 0.02$ and $c = 0.47 \pm 0.03$, or  $\alpha = 1.18 \pm 0.05$ and $\beta = -1.42 \pm 0.14$ with a scatter of 0.017  dex. As we can see from Table~\ref{tab1}, the values agree with both the re-analyses done for the CLASH sample and also with the Three Hundred Project values for both the temperatures in $(50-500) h^{-1}$ kpc range at a redshift of 0.07.


\subsection{Recap of  Results}
A tabular summary of our results for both the temperatures along with previous results from literature are collated in Table~\ref{tab1}. Therefore, using both $T_{ew}$ and $T_{mw}$, we get a FP with a very tight scatter  of about 0.017-0.019 dex. The best-fit FP value for $\beta$  using $T_{ew}$ and $T_{mw}$ is  in-between the self-similar and virial expectation values, whereas it is closer to the virial expectation for $\alpha$.
Our X-COP results using both the  temperatures are consistent with  the re-analysis of the CLASH sample carried out in G21 using only clusters which have temperatures less than 12 keV. For $T_{mw}$ they are also consistent with the aforementioned re-analysis done using only the relaxed cluster subset. The results using the mass weighted temperatures are consistent with the mass weighted results of G21 at a redshift 0.07 in $(50-500)h^{-1}\mathrm{\; kpc}$ range.
The cluster which is the farthest from the best-fit plane is A2319, which has the  highest temperature among all the clusters in the  sample. ~\citet{XCOPmain} classify this cluster as dynamically active based on the  central entropy estimates.
However, the X-COP clusters are on average relaxed systems due to the selection applied~\cite{XCOPmain}, and this has been  confirmed from the observed hydrostatic bias and the negligible amount of non-thermal pressure support~\cite{Eckertbias}. (See also ~\citet{Ettori2021}, which reaffirmed the earlier conclusions  using a new method to estimate the hydrostatic bias and non-thermal pressure.)

Because of the strong degeneracy between $\alpha$ and $\beta$ (cf. Fig.~\ref{fig:emcee_X-COP}), we check if the scatter changes by fixing the FP values to  the physically motivated virial or self-similar solutions.
The scatter of the X-COP clusters about the virial equilibrium plane ($\alpha$,$\beta$)= ($1,-1$), is found to be  0.036 dex, whereas the scatter about the plane corresponding to self-similar solution($\alpha$,$\beta$)= ($1.5,-2$), is equal to 0.027 dex using the error weighted temperature. Therefore, the scatter for the FP values corresponding to the virial as well as the self-similar solution is much larger than our best-fit values, which is in-between the two. \rthis{Note however that we can see from Fig.~\ref{fig:emcee_X-COP} that due to the strong anti-correlation between $\alpha$ and $\beta$, the FP values are  consistent with the virial expectations within 2$\sigma$.}

\begin{figure}
    \centering
    \includegraphics[width=1\columnwidth]{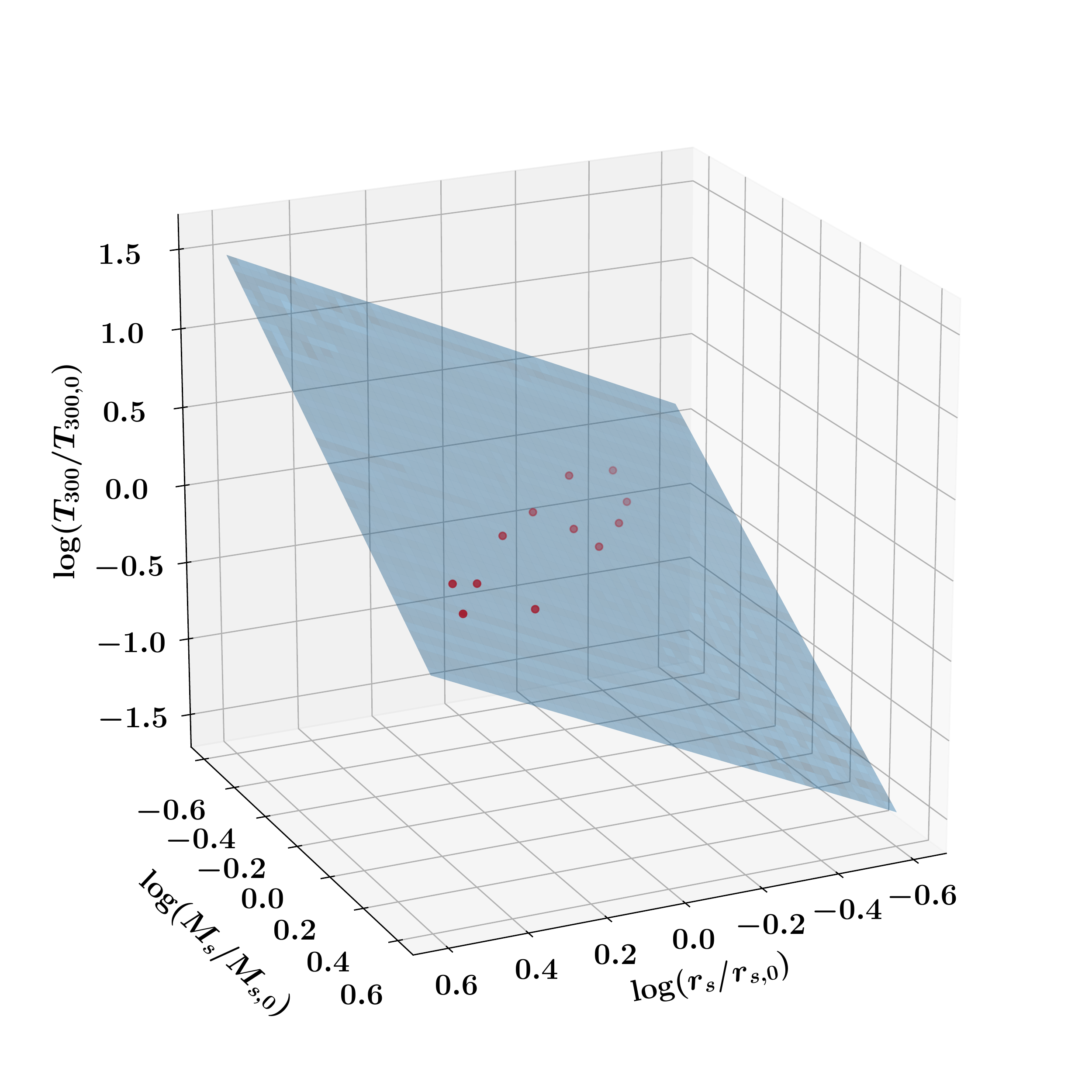}
    \includegraphics[width=1\columnwidth]{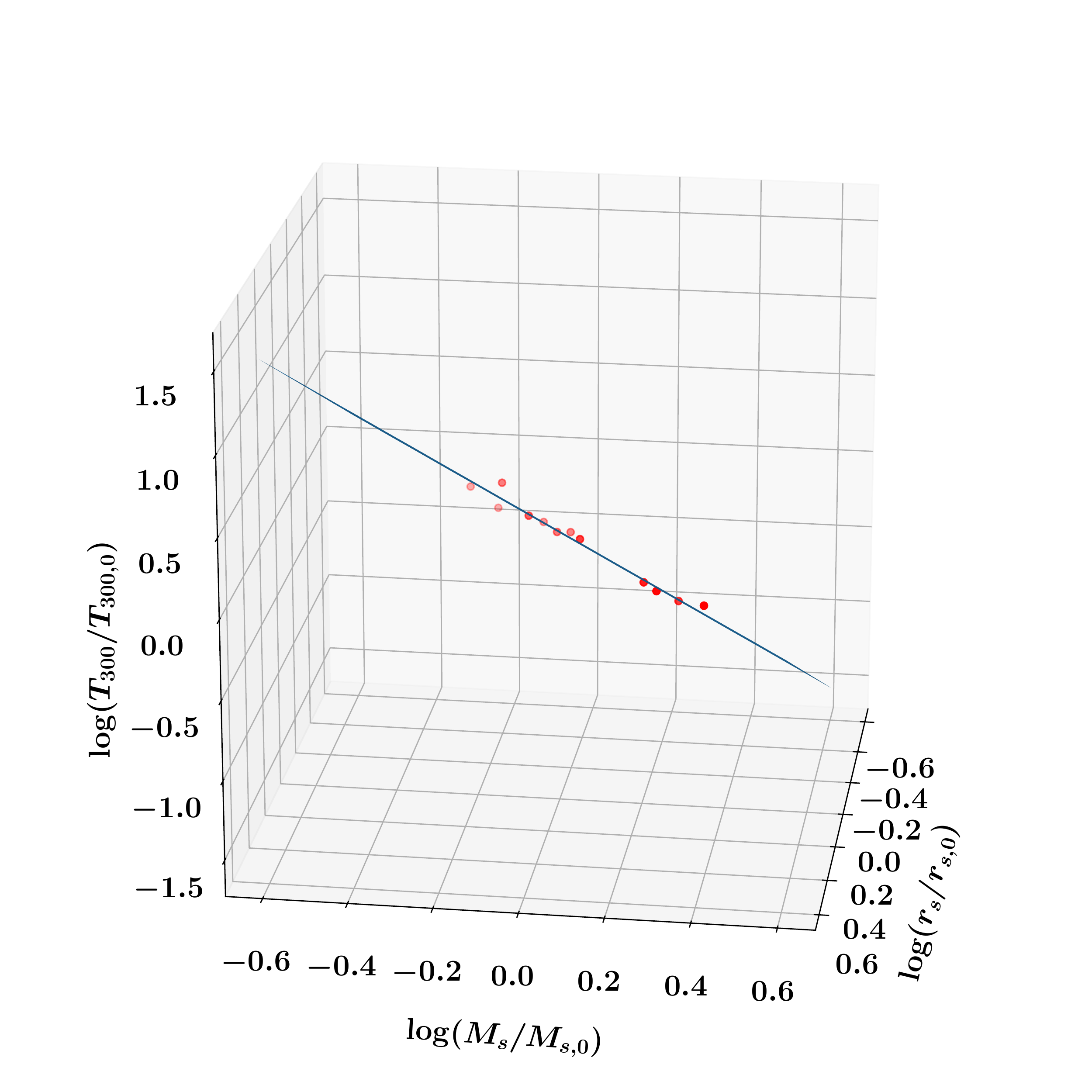}
    \caption{The X-COP data points (red filled circles) in $\log$ space of $(r_s/r_{s,0}, M_s/M_{s, 0}, T_{ew}/T_{ew,0})$ and the 
    best-fit plane to the data from different viewing angles. The equation of plane is $0.67x-0.54y+0.51z=0$. Note that the uncertainties of the points are typically about 10\% along all the axes and are not shown  in this figure.}
    \label{fig:fp}
\end{figure}

\begin{table*}[h]
    \centering
    \begin{tabular}{|l|c|c|c|c|}
     \hline
    \textbf{Dataset Used} & \boldmath$\alpha$ & \boldmath$\beta$    & \textbf{Scatter (dex)} &  \textbf{Reference}  \\ \hline
          X-COP ($T_{ew}$) & $1.06\pm 0.02$ & $-1.31\pm 0.12$ & 0.019  & This work\\
          X-COP ($T_{mw}$) & $1.18 \pm 0.05$ & $-1.42 \pm 0.14$ & 0.017 & This work\\
          CLASH (all clusters) & $1.8 \pm 0.5$ & $-2.3 \pm 0.7$ & 0.045 &  F18 \cite{Fujita2018}\\
          CLASH (relaxed clusters) & $1.8^{+0.8}_{-0.6}$ & $-2.4^{+1.0}_{-1.3}$& 0.029 &  G21~\cite{Garcia2021}\\
          CLASH ($T_X < 12 \mathrm{\;keV}$) & $1.0^{+0.6}_{-0.3}$ & $-1.3^{+0.4}_{-0.8}$ & 0.032 &  G21~\cite{Garcia2021}\\
          MUSIC & $1.30$ & $ -1.57$ & 0.025 & F18  \cite{Fujita2018} \\
          Ettori et al~\cite{Ettori10} & $1.15^{+0.29}_{-0.39}$&$ -1.54^{+0.51}_{-0.68}$ & 0.039 & \citet{Fujita2}\\
          Three Hundred Project using $T_{mw}(500h^{-1}\mathrm{kpc})^\dag$ & $1.18\pm 0.03$ & $-1.30 \pm 0.04$& 0.017 &  G21\cite{Garcia2021}\\
          Three Hundred Project using $T_{sl}(500h^{-1}\mathrm{kpc})^\dag$ & $1.23\pm 0.05$ & $-1.47\pm 0.07$& 0.022&  G21\cite{Garcia2021}\\
          \hline
    \end{tabular}
    \caption{Summary of fundamental plane analyses in literature including this work. Our best-fit values are in-between the virial ($\alpha=1$, $\beta=-1$) and self-similar model expectations ($\alpha=1.5$, $\beta=-2$). \\
    $^\dag$ The values mentioned here for G21 are at $z=0.07$. $T_{mw}$ stands for mass weighted temperature and $T_{sl}$ stands for spectroscopic like temperature. For values of $\alpha$ and $\beta$ obtained at other radii and at other redshifts, we refer readers to G21 \cite{Garcia2021}.}
    \label{tab1}
\end{table*}

\begin{table}[H]
    \centering
    \begin{tabular}{|c|c|c|}
        \hline
        \multicolumn{1}{|c|}{FP Parameters} & \multicolumn{1}{c|}{Pearson $r$} & \multicolumn{1}{c|}{$p-$value} \\ \hline
        $(a, b)$               & $-0.2$                           & $5.9\times 10^{-12}$           \\
        $(b, c)$               & $\enspace \; 0.7$                            & $6.7\times 10^{-148}$          \\
        $(c, a)$               & $-0.8$                           & $3.1\times 10^{-275}$          \\
        $(\alpha, \beta)$      & $-0.8$                           & $0$                            \\ \hline
    \end{tabular}
    \caption{Pearson $r$ coefficients and respective $p-$values for $a, b$ and $c$ obtained in Sec.~\ref{sec:t300} for the synthetic datasets (discussed in Sect.~\ref{sec:results} using the  error weighted temperature ($T_{ew}$) are tabulated in the first three rows. The final row corresponds to the Pearson $r$ coefficient obtained for $\alpha$ and $\beta$ estimated in Appendix~\ref{apx:A} from  Bayesian regression analysis using the MCMC chains. Note that the null hypothesis used to calculate the $p$-value is that the two datasets are uncorrelated.}
    \label{tab:pearsonr}
\end{table}

\section{Conclusions}
\label{sec:conclusions}
Recently, there have been multiple works 
with both data  and mock samples, which showed that galaxy  clusters form a tight FP in logarithmic space formed by the cluster temperature ($T_X$) halo mass ($M_s$), and scale radius ($r_s$), i.e. $T_X \propto M_s^{\alpha} r_s^{\beta}$. The first such study by F18 showed  using the CLASH cluster sample as well as an X-ray selected sample that $T_X \sim M_s^{1.5} r_s^{-2}$, in accord with self-similar infall model~\cite{Bertschinger85}, and deviates from the virial equilibrium solution given by $T_X \propto M_s r_s^{-1}$.
However, these results could not be confirmed in a recent study by G21 using a simulated cluster sample, who found that the FP is sensitive to the definition of the temperature adopted  as well as the cluster redshift. In their study, they concluded that the FP for mass-weighted temperatures is closer to the virial solution, whereas the FP for spectroscopic temperatures lies in-between the virial and self-similar solution. When the gas temperatures within $500 h^{-1}$ were used, the FP shifts towards the  self-similar solution. \rthis{Overall, they find that the CFP parameters obey the following constraints: $1<\alpha<1.5$ and  $-2<\beta<-1$. They also found that the CFP parameters have a mild dependence on the redshift and cluster dynamical state.}

Therefore, given these somewhat conflicting results,  we carry out one more test of the FP using the X-COP cluster sample consisting of 12 clusters selected from the Planck SZ survey. We use the $M_s$, and $r_s$ values from the NFW fits obtained by the X-COP team. We used two different temperatures for our analyses: the error-weighted  mean temperature in $(50-500)h^{-1}$   kpc as well as the mass-weighted temperatures integrated from to $(50-500)h^{-1}$. The FP analysis was done using PCA, similar to that in G18 and F21.
For $T_{ew}$, we also crosschecked our PCA-based results using Bayesian regression analysis. Our results  are tabulated in Table~\ref{tab1}. We obtain a scatter of 0.019 and 0.017 dex for $T_{ew}$ and $T_{mw}$, respectively. This is comparable to the scatters obtained from previous works.
We find that the best-fit values of $\beta$ is in-between the self-similar and virial solutions, whereas the value of $\alpha$ is closer to the virial value for $T_{ew}$, but slightly shifts towards larger values for $T_{mw}$. \rthis{Our FP values are therefore within the same ballpark found in G21. They are consistent with virial equilibrium within 2$\sigma$.}
Furthermore, we find in accord with the G21 results that  there is a mild dependence of the FP best-fit parameters  to the choice of the temperature definition.
Our best-fit FP values for the X-COP sample for both the temperatures are also consistent with the re-analyses of the CLASH sample done in G21, after excluding the clusters with $T_X >$ 12 keV.  Note however that the X-COP sample  also contains only clusters with $T_X < 12$ keV, which is a possible reason for the agreement
with this  re-analysis of a subset of the CLASH sample, done in G21. The results using $T_{mw}$
also agree with the re-analysis done in G21 using only the relaxed subset of the CLASH sample.
The best-fit FP values (using $T_{mw}$) are also consistent (within $1\sigma$) with the values obtained for the Three Hundred project simulated cluster sample using temperatures measured in $(50-500) h^{-1}$ kpc at a redshift of 0.07. \rthis {We also find in agreement with G21 that $\alpha$ and $\beta$ are strongly anti-correlated. Possible reasons for this will be explored in a future work.}

More detailed tests  of the fundamental plane along with its evolution as a function of the redshift and the cluster dynamical state should soon be possible with galaxy clusters  observed using the  CHEX-MATE sample~\cite{chexmate}.

\begin{acknowledgements}
We are grateful to Yutaka Fujita for a stimulating talk,  which motivated this work, and also for useful feedback on the manuscript. We also acknowledge the anonymous referee for  constructive feedback on the manuscript.
\end{acknowledgements}
\bibliography{references}

\appendix
\section{Fundamental Plane using Regression Analysis}
\label{apx:A}
As a cross-check of our PCA based results, we also test the FP using a power-law regression analysis between $T_X$, $M_s$, and $r_s$. We did this test with the error-weighted temperature in $(50-500)h^{-1}$ kpc.
Eq.~\ref{eq:FP} can be rewritten as:
\begin{equation}
    \log_{10} T_{ew} = \alpha \log_{10} M_s + \beta \log_{10} r_s + \delta
    \label{eq:FPeqn}
\end{equation}
with $\alpha=-b/c, \beta = -a/c$, and $\delta = d/c$. 
We now define $x', y'$ and $z'$ (using the same notation as G21) as:
\begin{equation}
\begin{aligned}
    x' &\equiv \log_{10} (M_s/M_{s,0})\\
    y' &\equiv \log_{10} (r_s/r_{s, 0})\\
    z' &\equiv \log_{10} (T_{ew}/T_{ew, 0})
\end{aligned}
\end{equation}
where the subscript 0 stands for the mean values. 
Now, the distance of a point $(x'_i, y'_i, z'_i)$ to this plane can be given as:
\begin{equation}
    d_i = \frac{| \alpha x_i' + \beta y'_i -z'_i + \delta| }{\sqrt{\alpha^2 + \beta^2 + 1}}
 \end{equation}

\begin{figure*}[h]
    \centering
    \includegraphics[width=1.7\columnwidth]{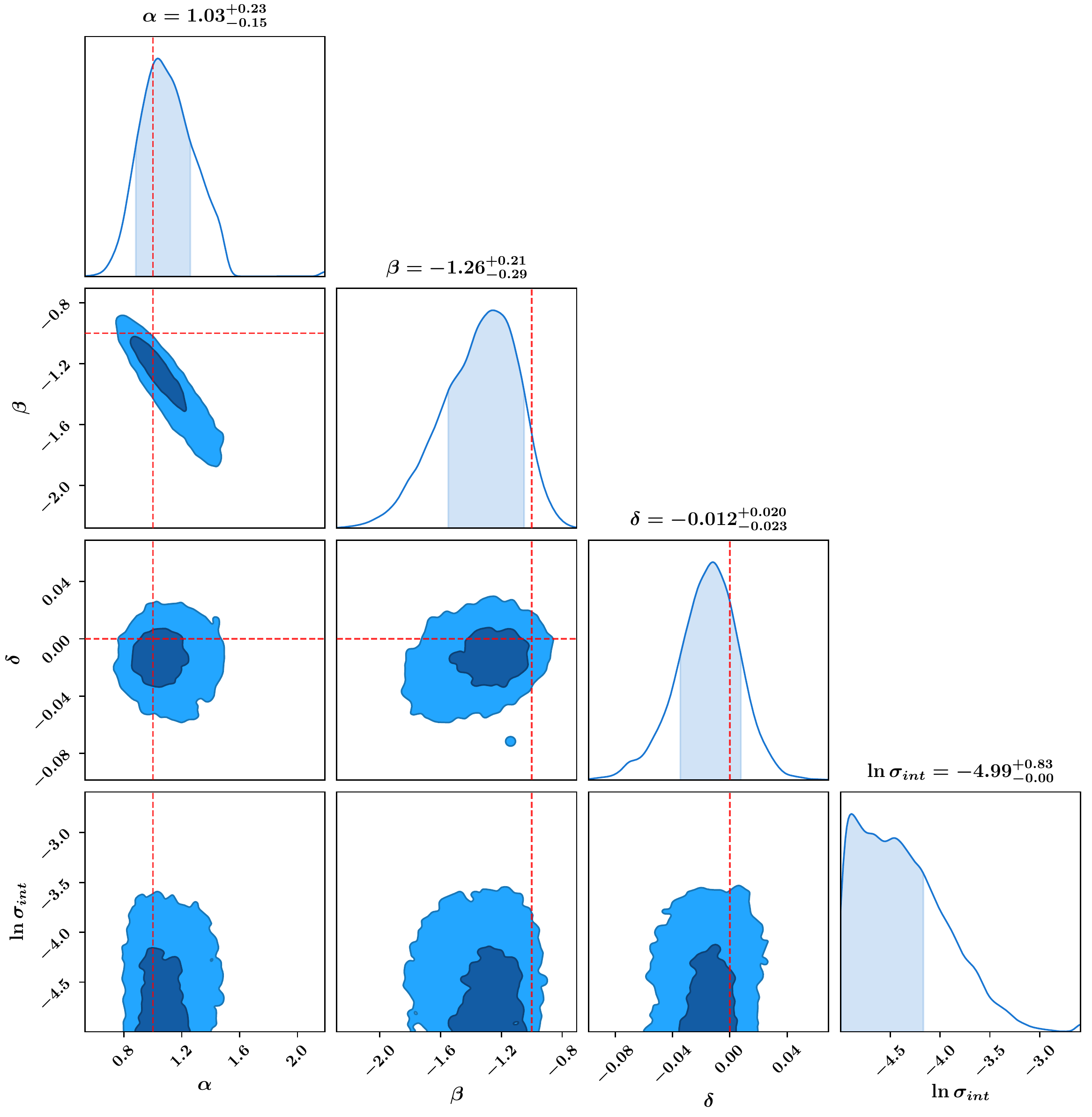}
    \caption{68\% and 95\% marginalised credible intervals obtained on fitting a plane of the form $z'=\alpha x' + \beta y' + \delta$ to X-COP galaxy clusters using \texttt{emcee}. Here, $\sigma_{\mathrm{int}}$ is the intrinsic scatter of the data about the best fit plane. This figure is generated with \textit{Chainconsumer} post-processing tool~\cite{Chainconsumer}.\rthis{ The dashed red line shows the virial equilibrium values for $\alpha$ and $\beta$, corresponding to $\alpha=1$ and $\beta=-1$. The FP values are consistent with the virial expectation values within $2\sigma$.}}
    \label{fig:emcee_X-COP}
\end{figure*}

Similar to the analysis done in G21, a plane can be fit to the data by minimizing the sum of square of distance of the points to the plane by defining a likelihood ($L$) as
\begin{equation}
    -2 \ln L = \sum_i (2 \pi \sigma_i^2) + \sum_i (d_i/\sigma_i)^2, 
    \label{eq:likelihood}
\end{equation}
where $\sigma_i^2 = \alpha^2 \sigma_{x'}^2 + \beta^2 \sigma_{y'}^2 + \sigma_{z'}^2 + \sigma_{\mathrm{int}}^2$. Here, $\sigma_{\mathrm{int}}$ is the intrinsic scatter which is also taken as a free parameter , similar to our previous works on RAR and tests of constant halo surface density~\cite{Pradyumna20,Gopika,Gopika21}
We then use \texttt{emcee} MCMC sampler\cite{emcee} to sample the likelihood,  using a flat priors on all the parameters with $0\le\alpha \le 5$, $-5 \le\beta \le 0$, $-0.1 \le\delta \le 0.1$ and $-5 \le\ln \sigma_{\mathrm{int}} \le 0$.
We use 20 walkers for MCMC analysis with a total number of steps being 6000 and the number of steps in burn-in being 500.
We obtain $\alpha = 1.03^{+0.23}_{-0.15}$, $\beta = -1.26^{+0.21}_{-0.29}$, $\delta = -0.0012^{+0.020}_{-0.023}$ and $\ln \sigma_{\mathrm{int}} = -4.99^{+0.83}_{-0.00}$. Therefore the best-fit values agree with the PCA-based estimates. It should be noted that the intrinsic scatter is bound by -5 as a lower prior and this is where the posterior gets  maximized using the  MCMC technique.
Fig.~\ref{fig:emcee_X-COP} shows 68\% and 95\% credible intervals obtained on marginalizing a posterior distribution maximized using \texttt{emcee}. We can see from Fig.~\ref{fig:emcee_X-COP} there is a strong degeneracy between $\alpha$ and $\beta$, and that $\alpha$ is anti-correlated with $\beta$. 

\end{document}